\documentstyle[a4,11pt,lscape]{article}

\title{GALAXY FORMATION --- A CONDENSATION PROCESS JUST AFTER
       RECOMBINATION}
\author{\sc G. Lessner\\
            FB Physik, Universität Paderborn, 33098 Paderborn\\
            Germany}
\date{\vspace*{2cm}}

\newcommand{\<}{\stackrel{\textstyle <}{\sim}}
\renewcommand{\>}{\raisebox{-.7ex}{$\stackrel{\textstyle >}{\sim}$}}

\begin{document}
\maketitle
\addtocounter{page}{-1}
\thispagestyle{empty}

\begin{abstract}
{\large\noindent A scenario of galaxy formation is put forward which is a process
of sudden
condensation just after recombination. It is essentially based on the fact
that the cosmic matter gas after recombination is a general relativistic
Boltzmann gas which runs within a few $10^6$ years into a state very close
to collision--dominated equilibrium. The mass spectrum of axially symmetric
condensation "drops" extends from the lower limit $M \simeq 10^5 M_\odot$
to the upper limit $M \simeq 10^{12} M_\odot$.
The lower limit masses are spheres whereas the upper limit masses are
extremely thin  pancakes. These pancakes contract
within a time of about $2,5 \cdot 10^9 y$ to fastly rotating spiral galaxies
with ordinary proportions. In this final state they have a redshift $z \simeq 3$.
At an earlier time during their contraction they are higly active and are
observed with a redshift $z \simeq 5$.\\

}
\end{abstract}
\newpage

1. \underline{{\bf Introduction}}\\
The formation of galaxies and large--scale structures in the universe is one of the
unsolved problems in the standard model given by the
Friedmann--Robertson--Walker--models. Although these models are altogether very
successful and
therefore widely accepted no way has been found hitherto to explain in this
framework the formation of the observed structures in a satisfactory manner.

The essential reason for this lack is the extreme smothness of the young universe.
This smoothness has been verified by the COBE data of the cosmic background
radiation. Of course, COBE observed anisotropies $\Delta T/T \simeq 10^{-5}$
of the background radiation only on a rather large scale of $7^0$ to $10^0$.
Hence one might observe on a much smaller scale much larger anisotropies.
Nevertheless this would be rather surprising. So it seems that one has to resign
to a very smooth universe when it was young. Carrying over the anisotropy
$\Delta T/T \simeq 10^{-5}$ of the background radiation to the density contrast
$\delta \varrho/\varrho  \simeq 10^{-5}$ at the end of the recombination period one is
confronted with the problem that these contrasts can not grow up to galaxies by
gravitational instability in a sufficiently small time [1]. Indeed, today galaxies
are observed with redshifts $z \simeq 5$ which corresponds to about $10^9 y$
after the big bang. A further problem almost never mentioned is
the fact that the galaxies with upper limit masses $(M \simeq 10^{12} M_\odot)$
are mainly flat and fastly rotating discs whereas the objects at the lower limit
$(M \simeq 10^5 M_\odot)$ are spheres.

Today it seems to be proved that the overwhelming part of matter in the universe
is non luminous dark matter. This is confirmed by the flat rotation curves in
galaxies as well as by gravitational lensing of galaxy clusters. Many cosmologists
believe that this dark matter is non baryonic exotic cold dark matter which is
subject only to gravitational interaction. They propose a scenario of structure
formation which is first of all structure formation in the cold dark matter, the
luminous baryomic matter then adds to these structures. The extremely cumbersome
numerical calculations yield interesting structures which look very much like to
those observed in the universe [2]. However, there are at least two problems in
this scenario. First of all, which particles form the cold dark matter? Hitherto
there is no experimental evidence for such exotic particles. Secondly, the
scenario might perhaps explain the very large--scale structures, it seems, however,
not to explain the origin of the single objects --- namely the galaxies and the
smaller objects down to the globular clusters.
 
\markright{\centerline{-- \arabic{page} --}\hspace*{30cm}}

In the following paper we propose a scenario of galaxy formation in an universe
without cold dark matter, that means, the dark matter is purely baryonic. A first
version of this scenario has been published by the author in two previous paper
[3,4]. This version describes only the \underline{final state} of galaxy
formation, not, however, the \underline{process}. This process has now been understood,
and thus the complete scenario can be represented.

The basic idea of the scenario is the fact that the cosmic matter gas is from the
end of the recombination period a general relativistic Boltzmann gas which runs
within a few $10^6$ years into a state very close to collision--dominated equilibrium.
Hence we sum up
briefly from the previous papers [3,4] in sections 2 and 3 the properties of the
cosmic matter gas after the recombination period and the properties of a general
relativistic Boltzmann gas in collision--dominated
equilibrium. In section 4 we describe the process of
galaxy formation which is a process of sudden condensation. In section 5 we
investigate the time evolution of the condensation "'drops"'. Finally we discuss
our results in a conclusion.

\bigskip

2. \underline{{\bf The cosmic matter gas after recombination}}

When the universe has cooled down to $T \simeq 5 \cdot 10^3 K$ the recombination
of the cosmic plasma to neutral H-- and He--atoms begins. It ends at
$$T \simeq 3 \cdot 10^3 K = T_{rek}  \qquad \qquad (1)$$
when all matter consists of neutral atoms. This recombination period is throughout
a drastic event in the history of the universe. One of the important consequences
is the drastic change of the range of interaction between the particles. It decreases
from the long range Coulomb interaction between charged particles before recombination
to the very short range collision interaction between neutral atoms after
recombination. In an earlier paper [5] the author argued that the range $r_c$ of
collision interaction in the cosmic matter consisting of 70 $\%$ hydrogen and 30 $\%$
helium is approximately given by
$$r_c \simeq 7 a_0 \qquad \qquad  (2)$$
with Bohr's radius $a_0$. Hence we have after recombination the relation
$$\stackrel{\textstyle r_c n^{1/3}}{\mbox{~}}\, \< 3 \cdot  10^{-7} \qquad\qquad \quad(3a)$$
with the particle density
$$n = n_H + n_{He} = 0,8 \frac{\varrho}{m_p} \qquad\qquad (3b)$$
($\varrho$ = mass density, $m_p$ = proton mass). From (3a) we can conclude that
the cosmic matter after recombination is a Boltzmann gas. Futhermore the author
calculated in [5] the ratio $t_c/t_{exp}$ where $t_c$ is the mean time between two
collisions and $t_{exp}$ is the characteristic expansion time of the universe.
This ratio takes the value
$$\frac{t_c}{t_{exp}} \simeq 4 \cdot 10^{-8} \left(\frac{T_{rek}}{T}\right)^2,
  T \le T_{rek} \qquad\qquad (4)$$
From this ratio we conclude that the cosmic matter runs from the end of
recombination $(T = T_{rek})$ into a state which is \underline{very close to
equilibrium}. But how does this process run and how much time takes it? To
answer this question we consider in the following section first of all the final
state of this process, namely the general relativistic Boltzmann gas in
collision--dominated equilibrium.\\

\markright{\centerline{-- \arabic{page} --}\hspace*{30cm}}

3. \underline{{\bf The general relativistic Boltzmann gas in collision--dominated
   equilibrium}}

We note that the cosmic matter after recombination runs into a state \underline{very
close} to equilibrium which is \underline{not exact} the equilibrium. Clearly,
the expansion of the universe does not admit an \underline{exact} equilibrium but
only a state \underline{very close} to equilibrium. Later on we shall assume that
the physics of this state is essentially that of exact equilibrium. Thus we deal
in this section with a general relativistic Boltzmann gas in collision--dominated
equilibrium. For
simplicity we assume a one--component gas with particle mass m.

We adopt the essential properties of a general relativistic Boltzmann gas in
collision--dominated
equilibrium from the comprehensive reviews given by Stewart [6] and Israel [7].
A space--time filled by such a gas has a time--like killing field
$$\beta^i = \frac{u^i}{kT} \qquad \qquad (5)$$
where the $u^i$ are the components of the 4--velocity and k is Boltzmann's constant
(latin indices running from 1 to 4, signature +++-). A reference frame ${\cal K}_0$
can be found such that
$$\beta^i = (0,0,0, const) \qquad \qquad (6)$$
In ${\cal K}_0$  the matter is in rest and the following relations hold
$$g_{ik} = g_{ik} (x^1, x^2, x^3),\quad T = T (x^1, x^2, x^3)
   \qquad \qquad\qquad (7)$$
$$g_{44} T^2 = const \qquad \qquad \qquad \qquad \quad\qquad(8)$$
$$m n (x^1, x^2, x^3) = \varrho (x^1, x^2, x^3) = const \cdot
  T \cdot K_2 (\frac{mc^2}{kT}) \qquad \qquad (9) $$

where n is the particle density and $K_2$ the modified Bessel function. The
specific relation (9) between the spatial changes of density and temperature is
due to the fact that in a general relativistic Boltzmann gas in
collision--dominated equilibrium the
chemical potential $\mu$ divided by T is \underline{spatially constant}.
We note explicitly that (9) is \underline{not} an \underline{equation of state}.
It describes \underline{exclusively} the \underline{spatial changes of density
and temperature} in an \underline{equilibrium solution} of Boltzmann's equation.

In the range of low temperatures $kT << mc^2$ an asymptotic expansion of $K_2$
yields
$$ \varrho = const \cdot T^{3/2} \cdot \exp [ - \frac{mc^2}{kT}] \qquad \qquad (10)$$
This extremely sensitive exponential relation between the spatial changes of
density and temperature has been found also by Dehnen and Obregon [8] in a quite
different way.

\markright{\centerline{-- \arabic{page} --}\hspace*{30cm}}

The metric is determind by the field equations. We consider small anisotropies of
the temperature, that means

\centerline{
~\qquad \parbox{6cm}{
\begin{eqnarray*}
 T & = & T_1 [1 + \tau (x^1, x^2, x^3)]\\
\\
 T_1 & = & const, \qquad | \tau (x^1, x^2, x^3) | << 1
\end{eqnarray*}}\qquad (11)}

Furthermore we assume the gravitational field to be weak in the sense

\begin{eqnarray*}
&&g_{ik} = \eta_{ik} + h_{ik} (x^1, x^2, x^3)\\
\\
&&\eta_{ik} = diag (1,1,1,-1) \qquad \qquad \qquad(12)\\
\\
&&| h_{ik} (x^1, x^2, x^3) | << 1
\end{eqnarray*}

Starting from eqs. (11) and (12) the author has given a detailed analysis of the
linearized field equations (with zero cosmological constant) in his previous
paper [4]. From this analysis follows the basic equation
$$
  \Delta \tau = -\kappa_0 \left\{ \frac{1}{2} \varrho_1 c^2\exp \left[
  - \frac{mc^2}{kT_1} (1 - \tau) \right] + a T_1^4 \right\} \qquad(13)
$$
where $\Delta$ is the Laplician, $\kappa_0 = 8 \pi G/c^4$ Einstein's gravitational
constant and $a = 7,564 \cdot 10^{-15} erg \cdot cm^{-3} \cdot K^{-4}$ the black--body
constant. Eq. (13) is for given background temperature $T_1$ and some start value
$\varrho_1$ of the density a partial differential equation for the temperature
anisotropies $\tau$ in a general relativistic Boltzmann gas in collision--dominated
equilibrium. The solution $\tau$ determines then the density by means of
$$
   \varrho = \varrho_1 \exp \left[ - \frac{mc^2}{kT_1} (1 - \tau) \right]
   \qquad\qquad(14)
$$
Eq. (13) says that in a general relativistic Boltzmann gas in collision--dominated
equilibrium anisotropies of the temperature \underline{must} exist. Indeed, with
$\tau \equiv 0$ it follows from eq. (13) that $\varrho_1 = 0$ and $T_1 = 0$ and
hence $\varrho \equiv 0$ and $T \equiv 0$. The existence of these spatial
fluctuations of temperature and density (by means of  eq. (14)) is an
\underline{intrinsically general relativistic effect}. It is due to the coupling
of the Boltzmann gas with the space--time metric via the field equations. In a
\underline{special relativistic} theory, however, space--time is fixed so that
 field equations do not exist. Then we have $g_{44} = const = -1$ and hence
$T = const$ and $\varrho = const$ according to eqs. (8) and (9).
Because of $mc^2/kT_1 << 1$ (later on we have $m = m_p$ and $T_1 \simeq 10^3 K$
so that $m_pc^2/kT_1 \simeq 10^{10} !$) it follows from eq. (14) that smallest
changes of $\tau$ lead to extremely drastic changes of $\varrho$. Hence
space--time is filled with sharply edged gas clouds although the temperature is
nearly isotropic. This is the final state in the condensation process described
later in section 4.2.

We are interested in a single gas cloud. Such a cloud is described by the
equations (see ref. [4])
$$\left.\parbox{9cm}{
$$
\tau_c \Delta y = - \frac{8 \pi G}{c^4} \left\{ {\frac{1}{2} \varrho_c c^2 \exp
\left[ - \frac{mc^2 \tau_c}{kT_1} (1 - y) \right] + a T_1^4 } \right\}
$$
$$ 
y = \frac{\tau}{\tau_c} \le 1, \qquad y_c = y (0,0,0) = 1
$$}\right\}\qquad (15 a)$$

$$
\varrho = \varrho_c \exp \left[ - \frac{mc^2 \tau_c}{k T_1} (1 - y) \right]
\qquad \qquad (15 b)
$$

Here the centre of the cloud is placed in the origin of the spatial coordinate
system, and $\varrho_c$ and $\tau_c$ are the values of $\varrho$ and $\tau$ in the
centre.

Again we emphasize that the density--temperature relation (10) or (15 b) holds in
the \underline{final quasi--equili-}
\underline{brium state}. The \underline{dynamical process}
ending in this final state will be treated below in section 4.2.

\markright{\centerline{-- \arabic{page} --}\hspace*{30cm}}

4. \underline{{\bf Galaxy formation}}\\
4.1 The cosmological background\\
Before describing in section 4.2 the galaxy formation as a sudden condensation
process we lay out the cosmological model in which this process runs. In
fig. 1 the age $t_u$ of the universe depending on the present day density $\varrho_0$
and the Hubble--parameter $H_0 = h \cdot km/sec Mpc$ is plotted where $k = 0$ marks
the flat universe. The underlying cosmological model is a standard model with zero
cosmological constant. 

The recent controversy about the Hubble--parameter between
Sandage/Tammann (small value of h) and Freedman et al. (large value of h) seems
to end in a mean value $h \simeq 70$. Furthermore the earlier large age of the
globular clusters of about $16 \cdot 10^9 y$ has been recently revised to
$(12 \pm 2) \cdot 10^9 y $ [9, 10]. So assuming a density $\varrho_{0, L} \simeq
2 \cdot 10^{- 31} g \cdot cm^{-3}$ of the luminous matter and taking into account the
dark matter by a factor 5 to 10 we have a present day density $\varrho_0 \simeq
10^{-30} g \cdot cm^{-3}$. From fig. 1 we then read an age of the universe between
$12 \cdot 10^9 y$ and $13 \cdot 10^9 y$ which is compatible with the age of the
globular clusters. So all is right. However, a flat universe as predicted by an
inflationary scenario in the very early universe seems to be ruled out. Obviously
we are living in an open universe.

\markright{\centerline{-- \arabic{page} --}\hspace*{30cm}}

The line element of an open standard model in quasi--cartesian coordinates reads
$$
ds^2 = \frac{1}{\left[ 1 - \frac{1}{4 R^2(t)}
 \sum\limits_\alpha (x^\alpha)^2 \right]^2}
\left[ (d x^{1})^2 + (d x^2)^2 + (d x^3)^2 \right] -
\underbrace{(d x^4)^2}_{c^2 dt^2} \qquad(16)$$

(greek indices running from 1 to 3). From the field equations with zero cosmological
constant we obtain then in the matter--dominated era $(T < T_{rek} = 3 \cdot 10^3 K)$
$$
ct = R \left\{ (1 + \frac{K}{R})^{\frac{1}{2}} - \frac{1}{2} \frac{K}{R} ln
\left[ \frac{(1 + K/R)^{\frac{1}{2}} + 1}{(1 + K/R)^{\frac{1}{2}}-1}\right]  \right\}
\qquad(17a)
$$

with
$$
K = \frac{1}{3} \kappa_0 c^2 \varrho_0 R_0^3, \quad R_0 = (H_0^2 - \frac{1}{3} \kappa_0
c^2 \varrho_0)^{- \frac{1}{2}} \qquad\qquad (17b)
$$

and where the index  $_0$ denotes the present day values. In the temperature range
$T \> 10^3 K$ equation (17a) can be well approximated by

$$
ct = \frac{1}{2K^\frac{1}{2}} R^{\frac{3}{2}} \qquad \qquad(18)
$$

From this equation we obtain
$$
R = (2c K^\frac{1}{2})^\frac{2}{3} t^\frac{2}{3} \qquad \qquad (19)
$$

and futhermore by use of $R \cdot T$ = const and eq. (17 b)
$$
t = \frac{1}{2c^2} \left ( \frac{3}{\kappa_0 \varrho_0} \right )^\frac{1}{2}
\left ( \frac{T_0}{T}\right )^\frac{3}{2} \qquad \qquad (20)
$$

with the present day temperature $T_0 = 2,7 K$ of the background radiation.

\bigskip

4.2 The condensation process

In section 2 we realized that the cosmic matter gas runs from the end of
recombination into a state very close to collision--dominated equilibrium. We
describe this state in a very good approximation by the exact equilibrium as
outlined in section 3. The single sharply edged gas clouds are then described
by eqs. (15) where we consider the cosmic matter as a one--component gas with
particle mass $m_p$. Below we shall find that the temperature in this state is
about $10^3 K$. Then we have even for $\tau_c \simeq 10^{-5}$ (COBE) in the
exponent of the density (15 b) the enormous factor
$$
\frac{m_pc^2 \tau_c}{kT_1} \simeq 10^5 \qquad \qquad \qquad(21)
$$

Hence, when $y$ drops only a very little below 1, say $y = 0,999,$
the density falls practically to zero.

Now we turn to the question how this process runs and how much time it takes.
Clearly, the general relativistic Boltzmann equation
$$
p^i \frac{\partial f}{\partial x^i} - \Gamma^i_{jk} p^j p^k \frac{\partial f}
{\partial p^i} = {\cal C} (f) \qquad (22)
$$

\markright{\centerline{-- \arabic{page} --}\hspace*{30cm}}

determines this process. There is no hope to solve this equation unless we make
some simplifications. Firstly we approcimate the collision term ${\cal C} (f)$
by the BGK--model [ 11 -- 19 ]
$$
{\cal C} (f) = - \frac{m_p}{t_c} (f - f_0) \qquad \qquad(23)
$$
where $t_c$ is the mean time between two collisions as in section 2 and $f_0$ is
the distribution in collision--dominated equilibrium. Secondly we calculate the
gravitational field $\Gamma ^i_{jk}$ only by means of the cosmological background
(16). This means

\centerline
{~\qquad\parbox{6cm}{\begin{eqnarray*}
&\Gamma^\gamma_{\alpha \beta} |_{x = 0} = 0\\
\\
&\Gamma^\alpha_{\beta 4}  = \frac{1}{c} \frac{\dot{R}}{R} \delta^\alpha_\beta, \quad
\Gamma^\alpha_{44}  =  0 \\
\\
&\Gamma^4_{\alpha \beta}  = \frac{1}{c} \dot{R} R \delta_{\alpha \beta}, \quad
\Gamma^4_{\alpha 4} = 0  =  \Gamma^4_{44}
\end{eqnarray*}}\qquad(24)}

where a dot denotes derivative with respect to t and $x = 0$ means
$x^1 = x^2 = x^3 = 0$. We suppose that at the end of recombination small
anisotropies of the background radiation are left from processes in the
very early universe and consider the regions where the above normalized
temperature variable $y$ is below 1. In these regions the later
(quasi)--equilibrium distribution $f_0$ approximately vanishes because the
particle density is practically zero (see above). Futhermore we assume that in
these regions
$$
f = f(t, p^1,p^2,p^3) \qquad \qquad (25)
$$
where the spatial momentum components $p^\alpha$ are related with the 4--th
component $p^4$ by the mass--shell condition
$$
g_{ij} p^i p^j = g_{\alpha \beta} p^\alpha p^\beta - (p^4)^2 = -m_p^2c^2
\qquad \qquad(26)
$$
In the cosmological epoch under consideration $(T \le 3 \cdot 10^3 K, m = m_p)$
we have
$$
g_{\alpha \beta} p^\alpha p^\beta << m^2_p c^2 \qquad \qquad(27)
$$
so that
$$
p^4 = m_p c \qquad \qquad(28)
$$
Making use of the property (25) we can place the points we consider at the
origin $x = 0$. Then the Boltzmann equation (22) reduces by means of the  collision
term (23) with $f_0$ = 0 and the relations (24), (25) and (28) to
$$
\frac{\partial f}{\partial t} - 2 \frac{\dot{R}}{R} p^\alpha \frac{\partial f}
{\partial p^\alpha} = - \frac{f}{t_c} \qquad \qquad(29)
$$

Next we integrate eq. (29) over the mass--shell (26) using the integration
element [6]
$$ \pi = \frac{1}{p^4} \sqrt{-g} dp^1 dp^2 dp^3 = \frac{1}{p^4} \sqrt{-g}
d^3 p \qquad(30)
$$

where $g = det g_{i j \cdot}$ By means of the metric (16) we have
$$
\pi |_{x = 0} = R^3 (t) \frac{d^3 p}{m_p c}\qquad \qquad (31)
$$
By use of
$$
\int f \pi |_{x = 0} = n(t) = \mbox{particle density} \qquad (32)
$$
the integration yields

\begin{eqnarray*}
\int \frac{\partial f}{\partial t} \pi |_{x = 0} 
 \displaystyle &=& \int \frac{\partial f}
{\partial t} R^3 (t) \frac{d^3 p}{m_p c}\\
 \displaystyle &=& \frac{d}{dt} \int f R^3 \frac{d^3 p}{m_p c} - 3 \int f R^2 \dot{R} \frac{d^3 p}
{m_p c}\\
 \displaystyle &=& \frac{d}{dt} \int f \pi |_{x = 0} - 3\frac{\dot{R}}{R} \int f \pi |_{x = 0}\\
 \displaystyle &=& \dot{n} - 3 \frac{\dot{R}}{R} n
\end{eqnarray*}

and
$$
\int p^\alpha \frac{\partial f}{\partial p^\alpha} \pi |_{x = 0} = \frac{R^3}
{m_p c} \underbrace{\int p^\alpha \frac{\partial f}{\partial p^\alpha} d^3 p}_
{(*)} = - 3 \int f \pi|_{x = 0} = - 3 n
$$

where we used in the integral (*) Gauß' theorem. Hence we arrive at
\markright{\centerline{-- \arabic{page} --}\hspace*{30cm}} 
$$
\dot{n} = - \left(\frac{1}{t_c} + 3 \frac{\dot{R}}{R}\right) n \qquad \qquad (33)
$$
The collision time $t_c$ is given by
$$
t_c = (n \cdot v_{th} \cdot \sigma)^{-1} \qquad \qquad(34a)
$$

with the termal velocity 
$$
v_{th} = \left (\frac{3kT}{m_p} \right )^{\frac{1}{2}} \qquad \qquad (34 b)
$$
and the cross section
$$
\sigma = \pi r_c^2 \qquad \qquad \qquad\quad(34c)
$$

$(r_c = 7a_0$ from eq. (2)). Futhermore it follows from eq. (19) that

$$
\frac{\dot{R}}{R} = \frac{2}{3t} \qquad \qquad (35)
$$
Hence eq. (33) takes the form
$$\parbox{9cm}{
$$
\dot{n} = -\left(\lambda_1 T^\frac{1}{2}n + \frac{2}{t}\right) n 
$$
$$
\lambda_1 = \pi r_c^2 \left ( \frac{3 k}{m_p} \right )^{\frac{1}{2}}
$$}(36)$$

 
From eq. (20 we derive)
$$
T = T_0 \left( \frac{3}{4 c^4 \kappa_0 \varrho_0} \right) ^\frac{1}{3}
\frac{1}{t^\frac{2}{3}} \qquad \qquad (37)
$$
so that we obtain finally the differential equation
$$\parbox{9cm}{
$$
\dot{n} = - \left (\frac{\lambda}{t^\frac{1}{3}} n + \frac{2}{t}\right )n 
$$
$$
\lambda = \pi r_c^2 \left( \frac{3 kT_o}{m_p} \right)^\frac{1}{2} \left(
\frac{3}{4 c^4 \kappa_0 \varrho_0} \right)^\frac{1}{6}
$$}(38)$$
for the particle density n between the gas clouds. This density decreases
on the one hand by collisions (first term in the paranthesis on the right
hand side of eq. (38)) and on the other hand by the cosmological expansion
(second term). Below we shall find that after recombination at first the
collision term is greater than the expansion term by about eight orders of
magnitude. However, this extreme superiority decreases very rapidly.

The process (38) starts at the end of recombination at $T = T_{rek} =
3 \cdot 10^3 K.$ Considering universes with
$$
\varrho_0 = \alpha \cdot 10^{30} gcm^{-3};\quad \alpha = 0.5,1,2.5 \qquad (39)
$$
we obtain from eq. (20) the initial time
$$
t = t_{rek} = \alpha^{-\frac{1}{2}} \cdot 0,57 \cdot 10^6 y \qquad \qquad(40)
$$

The particle density at $t = t_{rek}$ is given by $n_{rek} = \varrho_{rek}/m_p$
where 

$$
\varrho_{rek} = \alpha \cdot \bar{\varrho}_{rek}, \quad \bar{\varrho} = 1,4 \cdot
10^{-21} gcm^{-3} \qquad \qquad(41)$$

with $\bar{\varrho}_{rek}$ the density at the end of recombination in an
universe with $\varrho_0 = 10^{-30} gcm^{-3}$. Hence we have the initial particle
density
\markright{\centerline{-- \arabic{page} --}\hspace*{30cm}}
$$
n = n_{rek} = \alpha \cdot 8,2 \cdot 10^2 cm^{-3} \qquad \qquad(42)
$$
Next we write the differential equation (38) in a dimensionless form by
introducing
$$ 
t = x \cdot 10^6 y, \quad n = \alpha \cdot \mu \cdot 8,2 \cdot 10^2 cm^{- 3}
\qquad \qquad (43a)
$$

with the initial values
$$
x_{rek} = \alpha^{-\frac{1}{2}} \cdot 0,57, \quad \mu_{rek} = 1 \qquad \qquad(43b)
$$

Then eq. (38) reads
$$\parbox{9cm}{
$$ 
\mu' = \frac{d \mu}{dx} = - (\beta \frac{\mu}{x^\frac{1}{3}} + \frac{2}{x})
\mu 
$$

$$\beta = \alpha^\frac{5}{6}  \cdot 8,0 \cdot 10^7$$} (44)$$

The subsitution $u = \mu^{-1}$ leads to the linear differential equation
$$
u' = \frac{2}{x} u + \frac{\beta}{x^\frac{1}{3}} \qquad \qquad(45)
$$
with the general solution
$$
u (x) = C x^2 - \frac{3}{4} \beta x^{\frac{2}{3}} \qquad \qquad (46)
$$

The constant C is determined by the initial condition $\mu (x_{rek}) = 1$
Hence we arrive at
$$
\mu (x) = \left [ (x^{-2}_{rek} + \frac{3}{4} \beta x_{rek}^{-\frac{4}{3}}) x^2 -
\frac{3}{4} \beta x^{\frac{2}{3}} \right ]^{-1} \qquad \qquad (47)
$$

For $x$ well larger than $x_{rek}$ this solution can be written in the form
$$
\mu (x) = \frac{4}{3 \beta} \, \frac{1}{x^2 \left(x_{rek}^{-\frac{4}{3}} - x^{-\frac{4}{3}}\right)}
\qquad \qquad (48)
$$
The dimensionless form (44) of the differential equation (38) shows the
enormous collision rate (first term in the paranthesies on the right hand side)
just after recombination.

We suppose that the process of cloud formation is worked out when the
cosmological expansion rate predominaties more or less strongly the collision rate.
Hence we determine the end $x_1$ of the condensation process by
$$
\frac{2}{x_1} = \, b \ \beta \, \frac{\mu (x_1)}{x_1^{\frac{1}{3}}} \, , \,  b \simeq 10
\qquad \qquad (49)
$$
\markright{\centerline{-- \arabic{page} --}\hspace*{30cm}}

which by means of the solution (48) yields
$$
x_1 = \left(\frac{2}{3} b + 1 \right)^\frac{3}{4} x_{rek}\qquad \qquad(50)
$$
From eq. (50) we obtain the cosmological instant
$$
t_1 = \left( \frac{2}{3} b + 1 \right) ^\frac{3}{4} \alpha^{-\frac{1}{2}} \cdot 0,57
\cdot 10^6 y \qquad (51)
$$
of the end of the condensation process, its duration

$$
\Delta t = \left[\left(\frac{2}{3} b + 1 \right)^\frac{3}{4} - 1 \right] 0,57 \cdot \alpha^{-\frac{1}{2}}
\cdot 10^6 y \qquad(52)
$$

and by use of eq. (37) the temperature

$$
T_1 = T_{rek} \left( \frac{x_{rek}}{x_1} \right)^{\frac{2}{3}} = \left(\frac{2}{3} b + 1
\right)^{-\frac{1}{2}} T_{rek} \qquad(53)
$$

Finally we obtain from eq. (49) the value

$$
\mu_1 = \mu (x_1) = \frac{3,6 \cdot 10^{-8}}{b \left( \frac{2}{3}b + 1 \right)
^{\frac{1}{2}} \cdot \alpha^{\frac{1}{2}}} \qquad \qquad(54)
$$

In tables 1 and 2 we give for $b = 10$ and $b = 20$ the values of $t_1, \Delta t,
\mu_1$ and $T_1$ where $\alpha$ runs through the values according to eq. (39).

\markright{\centerline{-- \arabic{page} --}\hspace*{30cm}}

The last row in table 1 gives the ratio $\frac{1}{2} c \Delta t / R_1$ where
$R_1 = R (t_1)$ and $R (t_1)$ is calculated from eq. (19) by use of a mean
Hubble--parameter $h = 70.$ In section 4.3 we shall consider axially symmetric
gas clouds.
The maximal linear extension of these clouds is $c \Delta t$ because the clouds
must be causally connected. From the centre of the clouds to the edge we have then
a maximal extension $\frac{1}{2} c \Delta t$. This length must be much smaller
than $R_1$ otherwise the weak field expansion (12) does not hold (see the line
element (16)).

4.3 The mass spectrum\\
We can be sure that protogalaxies had been axially symmetric objects with their
lower limit of spherically symmetric globular clusters. Hence we are interested
in axially and spherically symmetric solutions of the single--cloud equations
(15).

We fix the particle mass as above by $m = m_p$ and let the anisotropies $\tau_c$
of the temperature vary in the range
$$ 
10^{-4} \ge \tau_c \ge 10^{-7} \qquad \qquad (55)
$$

The background temperature $T_1$ is determined by the parameter b (see tables
1 and 2). What about the density $\varrho_c$ in the centre? In section 4.2 we
established that the condensation process, that means the formation of the very
sharp edges of the gas clouds, runs very rapidly. Furthermore the linear
extension of the clouds is much smaller than the curvature radius $R$. Thus the
clouds are not subject to the cosmic expansion and hence we can assume
$$
\varrho_c = \varrho_{rek} = \alpha \bar{\varrho}_{rek} \qquad \qquad(56)
$$
\markright{\centerline{-- \arabic{page} --}\hspace*{30cm}}
with $\varrho_{rek}$ and $\bar{\varrho}_{rek}$ according to eq. (41). As in the
previous papersj [3,4] we introduce
$$
\frac{1}{l_{rek}^2} = \frac{8 \pi G}{c^2} \bar{\varrho}_{rek}, \quad l_{rek} =
6,6 \cdot 10^5 ly \qquad \qquad (57)
$$
and the dimensionless variable
$$
\bar{x}^\alpha = \frac{1}{\sqrt{\tau_c} l_{rek}} x^\alpha \qquad \qquad(58)
$$
Then eqs. (15) take the form

$$\left.\parbox{10cm}{
$$\Delta_{\bar{x}} y = - \alpha_1 \exp \left[ -\frac{m_pc^2 \tau_c}{kT_1}
(1 - y)\right] -q$$
$$y \le 1, \qquad y (0, 0, 0) = 1$$
$$ \alpha_1 = \frac{1}{2} \alpha, \qquad q = a T^4_1 /\bar{\varrho}_{rek} c^2 $$
}\right\} \qquad (59 a)$$

$$
\varrho = \alpha \bar{\varrho}_{rek} \exp \left[-\frac{m_pc^2 \tau_c}{kT_1} (1 - y) \right]
\qquad \qquad(59 b)
$$

where $\Delta_{\bar{x}}$ denotes the Laplacian with respect to $\bar{x}^\alpha$.\\

4.3.1 Spherically symmetric clouds

We denote the dinensionless radial coordinate by $x$ so that $ y = y (x).$
Then eq. (59a) reads
$$
y'' + \frac{2}{x} y' = - \alpha_1 \exp \left[-\frac{m_p c^2\tau_c}{kT_1}
(1 - y) \right] - q \qquad \qquad(60a)
$$
$$y (0) = 1 \qquad \qquad \qquad\qquad(60b)$$
To the natural boundary condition $y (0) = 1$ we add the regularity condition
$$ 
y'(0) = 0 \qquad \qquad\qquad(60c)
$$
In his previous paper [3] the author investigated the differential equation (60)
in detail. It turned out that the natural boundary condition (60b) implies the
regularity (60c). The solution takes in the very vincinity of the centre the
form

$$\parbox{9cm}{
$$
y_i (x) = 1 - \frac{\alpha_1 + q}{\beta} + \frac{\alpha_1 + q}{\beta^{\frac{3}{2}} x}
\sin (\beta^{\frac{1}{2}} x)$$

$$\beta = \frac{m_p c^2 \tau_c \alpha_1}{kT_1}$$}\qquad(61)$$

This solution is used as start for a numerical integration of eq. (60a) by a
Range-Kutta-method.

The mass M of the gas clouds in the centre of the anisotropies is given by
$$
M = 4 \pi \int^{r_1}_0 \varrho r^2 dr, \quad r = \sqrt{\tau_c} l_{rek} x 
\qquad\qquad(62)
$$
with the density $\varrho$ according to eq. (59 b). In section 4.2 we have shown
that the particle density at the boundary of the clouds drops to $\mu_1 n_{rek}$
when the condensation process is closed. Hence $r_1$ in the mass integral (62)
is strictly speaking given by $\varrho (r_1) = \mu_1 \varrho_c$ (see eq. (56)).
However, the numerical calculations show that the masses practically do not change
if we fix $r_1$ by $\varrho (r_1) = 10^{-5} \varrho_c$. This is due to the
extremely rapid decrease of $\varrho$.

Writing the mass formula (62) in a dimensionless form we obtain
\markright{\centerline{-- \arabic{page} --}\hspace*{30cm}} 

$$\parbox{9cm}{
$$
\frac{M}{M_\odot} = 2,13 \cdot 10^{18} \cdot \alpha \cdot \tau_c^\frac{3}{2}
\int^{x_1}_0 \exp [- \gamma (1 - y (x))] x^2 dx$$
$$
\gamma = \frac{m_p c^2 \tau c}{kT_1}, \quad \exp [- \gamma (1 - y (x_1))] 
= 10^{-5}$$}
\qquad(63)$$

The integral in eq. (63) can be easily evaluated by use of the numerical solution
$y (x)$. In tables 3 and 4 the results are listed where $D = 2r_1$ is the
diameter of the clouds. \underline{All} results are \underline{independent} of
the central value $\tau_c$ in the range (55). Finally we note that the radii
$\frac{1}{2} D$ of the clouds are very much smaller than $R_1$ (see the end of
section 4.2).\\

4.3.2 Axially symmetric clouds\\
We introduce dimensionless cylindrical coordinates $x$ (distance from the
symmetry axis) and $\zeta$ (z--coordinate) so that $y = y (x,\zeta)$. Then eq.
(59 a) reads
$$
\frac{\partial^2 y}{\partial x^2} + \frac{1}{x} \frac{\partial y}{\partial x}
+ \frac{\partial^2 y}{\partial \zeta^2} = - \alpha_{1} \exp [- \gamma (1 - y)]
- q \qquad(64 a)
$$

with $\gamma$ according to eq. (63). Again we have the natural boundary condition
$$
y (0,0) = 1 \qquad \qquad(64 b)
$$

and the regularity condition
$$
\frac{\partial y}{\partial x} (0, \zeta) = 0 = \frac{\partial y}{\partial \zeta}
(x, 0) \qquad \qquad \qquad (64c)
$$

Also the partial differential equation (64) has been investigated in detail by
the author in his previous paper [3]. The solution is composed of an interior
solution
$$
y_i (x, \zeta) = 1 - \frac{1}{2} (\alpha_1 + q) \left[\frac{1}{2} C x^2 + (1 - C)
\zeta^2 \right], \qquad(65 a)
$$

an exterior solution (when the density term on the right hand side of eq. (64 a)
has practically fallen to zero)
$$
y_a (x, \zeta) = 1 - \frac{1}{2} q \left[\frac{1}{2} C x^2 + (1 - C) \zeta^2 \right]
 \qquad(65 b)
$$

and a fit function
$$
y_f (x, \zeta) = 1 + \epsilon - \sqrt{2 \epsilon (\alpha_1 + q )} \sqrt{\frac{1}
{2} C x^2 + (1 - C) \zeta^2} , \qquad \epsilon << 1 \qquad(65 c)
$$

with a flattening parameter C in the range
$$
0 < C < 1 \qquad \qquad(65 d)
$$

The fit function connects the interior and exterior solution. It is fit 
smoothly to the
interior solution on a surface $y_i (x, \zeta) = 1 - \epsilon = $ const. i.e.
$$
\frac{1}{2} C x^2 + (1 - C) \zeta^2 = \frac{2 \epsilon}{\alpha_1 + q} \qquad\qquad(66)
$$

and intersects the exterior solution on the surface
$$
\frac{1}{2} C x^2 + (1 - C) \zeta^2 = \frac{2 \epsilon}{q^2} (\sqrt{\alpha_1 + q}
+ \sqrt{\alpha_1})^2 \qquad \qquad(67)
$$

The solution is spherically symmetric for $C = \frac{2}{3}$. Then the interior
solution (65 a) reads
$$
y_i (x, \zeta) = 1 - \frac{1}{6} (\alpha_1 + q) (x^2 + \zeta^2) \qquad(68)
$$

which agrees with the spherically symmetric solution (61) if one takes the first
two terms in the expansion of the sin.

On the surface of intersection (67) the density term on the right hand side
of eq. (64 a) has the value

$$
\alpha_1 \exp [-\gamma (1 - y_a)] = \alpha_1 \exp \left[- \frac{\gamma \epsilon}{q}
(\sqrt{\alpha_1 + q} + \sqrt{\alpha_1} )^2 \right] = \epsilon_1 \qquad(69)
$$
\markright{\centerline{-- \arabic{page} --}\hspace*{30cm}}

We express the parameter $\epsilon$ in terms of $\epsilon_1$ and obtain
$$
\epsilon = - \frac{q ln (\epsilon_1/ \alpha_1)}{\gamma (\sqrt{\alpha_1 + q} +
\sqrt{\alpha_1})^2} \qquad \qquad(70)
$$

Now we determine for given $T_1 (\leadsto q, \gamma)$ and $\alpha_1$ the
parameter $\epsilon$ by fixing $\epsilon_1$ such that the mass integral for
$C = \frac{2}{3}$ (spherical symmetry) agrees with the spherically symmetric
masses calculated in section 4.3.1. As above we have the dimensionless mass
formula

$$
\frac{M}{M}_\odot = 2,13 \cdot 10^{18} \cdot \alpha \cdot \tau_c^{\frac{3}{2}}
\int\limits^{x_1}_0 \int\limits^{\zeta_1}_0 \exp [- \gamma (1 - y (x, \zeta))]
xdxd \zeta \qquad (71a)
$$

where again at the upper limits of the integral the density has fallen to
$10^{-5} \varrho_c, i.e.$

$$
\exp [- \gamma (1-y (x_1, 0))] = 10^{-5} = \exp [-\gamma (1 - y (x, \zeta_1)) ] ,\qquad
x \le x_1 \qquad (71 b)
$$

and $y(x, \zeta)$ is the composed solution (65).

In tables 5 and 6 the values of $\epsilon_1$ are listed where we listed for
reasons of later comparison in the last row the values of $\mu_1$ (see eq. (54)).
Again the values of $\epsilon_1$ \underline{do not depend on $\tau_c$} (the values
of $\epsilon$, therefore, depend on $\tau_c$ since $\gamma$ depends on $\tau_c$).

For $C \rightarrow 0$ and $C \rightarrow 1$ the clouds are pancake-like and
cigar-like respectively (see fig. 2). We note that these configurations exist
in an exact equilibrium state in the global rest frame (see section 3),
especially the pancakes do \underline{not rotate}. They are hold by a complicated
balance of pressure and gravitational forces where the gravitational forces come
also from the radiation field [4].

What are the minimal and maximal value of C? We know from section 4.2 and eq.
(56) that at the end of the condensation process the density at the boundary
of the clouds drops to $\mu_1 \varrho_c$. Hence we obtain from eqs. (59 b),
(56) and (58) a relation between the flattening parameter C and the semiaxises
$r_0$ and $z_0$ by means of

$$\parbox{9cm}{
$$
\exp [- \gamma (1 - y (x_0, 0))] = \mu_1 = \exp [-\gamma (1 - y(0, \zeta_0))]
$$

$$
x_0 = \frac{r_0}{\sqrt{\tau_c}l_{rek}} , \qquad \zeta_0 = \frac{z_0}
{\sqrt{\tau_c} l_{rek}}
$$}\qquad(72)$$

From tables 5 and 6 we see that $\epsilon_1/\alpha_1 << \mu_1$. According to eq.
(69) this means that the exponential is on the surface of intersection between
the fit function and the exterior solution very much smaller than $\mu_1$.
Hence $y$ in eq. (72) lies in the domain of $y_f$. By use of eqs. (65 c) and (70)
we obtain then from eq. (72)
$$\left. \parbox{2.5cm}{
$$\frac{\displaystyle r_0 \sqrt{C}}{\displaystyle l_{rek}}$$
$$\frac{\displaystyle z_0 \sqrt{2(1 - C)}}{\displaystyle l_{rek}}$$
\hspace*{14.5cm}(73)}\right\} =
- 0,96 \cdot 10^{-5} (T_1 [10^3 K])^\frac{1}{2} \left\{ln \mu_1 +
\frac{qln(\epsilon_1 / \alpha_1)}{(\sqrt{\alpha_1 + q} + \sqrt{\alpha_1})^2} \right\}
\cdot \frac{\sqrt{\alpha_1 + q} + \sqrt{\alpha_1}}{[-q (\alpha_1 + q)ln
(\epsilon_1 / \alpha_1) ]^{\frac{1}{2}}}$$

Since the clouds must be causally connected at the end of the condensation process
the semiaxises $r_0$ for $C \rightarrow 0$ and $z_0$ for $C \rightarrow 1$
are limited by
$$
2 r_{0, max} = c \Delta t = 2 z_{0, max} \qquad  \qquad(74)
$$
\markright{\centerline{-- \arabic{page} --}\hspace*{30cm}}

where $\Delta t$ is the duration of the condensation process according to tables
1 and 2. With these upper limits of $r_0$ and $z_0$ we calculate from eq. (73)
the values $C_{min}$ and $C_{max}$. The results are listed in tables 7 and 8.

With the composed solution (65) we can now easily compute from the mass formula
(71) the masses of the clouds for given flattening parameters C where again it
is sufficient to break off the integration when the density drops to $10^{-5}
\varrho_c$ (instead of $\mu_1 \varrho_c$). The results are listed in tables 9 
and 10. Again all is independent of the central value $\tau_c$ in the range (55).

The mass spectrum we obtained is rather satisfactory. In fact, it shows the
observed upper limit masses $M \simeq 10^{12} M_\odot$ 
as well as the lower limit masses $M \simeq 10^5 M_\odot$ where, beyond it, the
lower limit masses are \underline{spheres}. However, the upper limit masses
$M \simeq 10^{12} M_\odot$ are extremely thin pancakes, and galaxies are not
observed in this form. Thus we investigate in the following section the evolution
of the condensation "'drops"' where we argue as in the author's previous paper
[3].\\

5)\underline{{\bf The evolution of the clouds after the condensation process}}\\
After the condensation "drops" have formed a few $10^6$ years after the big bang
the universe cools further down, that means, the background temperature $T_1$
decreases. If we assume that the quasi-equilibrium at first stands we conclude
from eq. (73) that the semiaxises $r_0$ and $z_0$ change by means of
$T_1^\frac{1}{2}$ in front of the curly bracket and $q \sim T_1^4$ (see eq.
(59a)). The second term in the curly bracket can be neglected against $ln \mu_1$
and hence we have approximately
$$
r_0 \sim \frac{1}{T_1^{\frac{3}{2}}}\, , \, z_0 \sim \frac{1}{T_1^{\frac{3}{2}}} \qquad\qquad(75)
$$

This means that the density $\varrho_{cl}$ in the clouds decreases as
$$
\varrho_{cl} \sim T_1^{\frac{9}{2}}\qquad\qquad \qquad \qquad (76)
$$
Then the characteristic expansion time of the clouds is given by

$$
\tau_{\exp} = | \frac{\varrho_{cl}}{\dot{\varrho}_{cl}} | = \frac{2}{9} |
\frac{T_1}{\dot{T_1}} | = \frac{2}{9} \frac{R}{\dot{R}} \qquad\qquad (77)
$$

where we used the cosmological law $R \cdot T = const.$ We compare this
characteristic time with the mean time $\tau_{coll}$ between two collisions
of particles in the clouds. It is given by
$$
\tau_{coll} = (n_{cl} \cdot v_{th} \cdot \sigma)^{-1} \qquad(78a)
$$
with the particle density
$$
n_{cl} = \frac{\varrho_{cl}}{m_p}\qquad \qquad  \qquad (78 b),
$$
the thermal velocity
$$
v_{th} = \left(\frac{3 kT_1}{m_p}\right)^\frac{1}{2}\qquad \qquad(78c)
$$
and the cross section
$$
\sigma = \pi r_c^2 \qquad \qquad (78 d)
$$
(see eq. (34)). According to eq. (76) the density $\varrho_{cl}$ can be written
in the form
$$
\varrho_{cl} = \varrho_{cl,1} \left(\frac{T_1}{10^3 K}\right)^\frac{9}{2} \qquad (79 a)
$$
where we use a mean value $T_1 = 10^3 K$ at the end of the condensation process and
hence
\markright{\centerline{-- \arabic{page} --}\hspace*{30cm}}
$$
\varrho_{cl,1} \simeq \alpha \cdot 10^{-21} g \cdot cm^{-3}\qquad \qquad (79 b)
$$
(see eq. (41)). In eq. (77) we use the field equation
$$
\frac{R}{\dot{R}} \simeq \frac{1}{c^2} \left( \frac{3}{\kappa_o \varrho} \right)^
\frac{1}{2} \qquad \qquad(80)
$$
with the mean cosmological density $\varrho$ which is correlated to the background
temperature $T_1$ by means of $\varrho \sim T_1^3$. Hence we have
$$
\varrho = \varrho_1 \left( \frac{T_1}{10^3 K} \right)^3 \qquad \qquad(81 a)
$$

with the cosmological density $\varrho_1$ at $T_1 = 10^3 K$ given by
$$
\varrho_1 = (\frac{1}{3})^3 \alpha \bar{\varrho}_{rek} = \alpha \cdot 5 \cdot
10^{-23} gcm^{-3} \qquad \qquad(81 b)
$$
Sticking all together we find
$$
\frac{\tau_{coll}}{\tau_{\exp}} \simeq \alpha^{-\frac{1}{2}} \cdot 2 \cdot 10^{-8}
\left(\frac{10^3 K}{T_1}\right)^{\frac{7}{2}} \qquad (82)
$$
This relation tells us that the clouds when expanding run rather rapidly out of
their quasi--equilibrium state so that the contraction due to the decreasing
temperature begins. We assume that the two effects nearly cancel so that the
clouds after the condensation process at first simply drift apart
until the universe has cooled down drastically, say down to $T \simeq 100 K
= T_2$. At this time the universe is about $10^8 y$ old and we assume that
the clouds begin to contract. 

We consider this contraction in more detail.
Especially we are interested in the upper limit masses $M \simeq 10^{12} M_\odot$
which are extremely thin pancakes. Indeed, with $C \simeq 10^{-7}$ (see tables 7
to 10) we obtain from eq. (73)
$$
\frac{z_o}{r_o} = \sqrt{\frac{C}{2(1 - C)}} \simeq \frac{1}{4500} \qquad \qquad (83)
$$
How do these clouds contract? To answer this question we compare the
gravitational energy $E_G$ of the clouds with their thermal energy $E_{th}$ after
they have drifted apart and the universe has cooled down to $T_2 \simeq 100 K$.
We have
$$
E_G \simeq - \frac{GM^2}{V^\frac{1}{3}} ,\quad E_{th} \simeq P \cdot V,\quad P \simeq
\frac{\bar{\varrho} k T_2}{m_p} \qquad\qquad(84 a)
$$
with the volume V of the clouds and a nearly constant density
$$
\bar{\varrho} \simeq \alpha \bar{\varrho}_{rek} \simeq \alpha \cdot 10^{-21}
gcm^{-3} \qquad \qquad(84 b)
$$
From these equations we obtain
$$
| E_G | \simeq \frac{Gm_p {\bar\varrho}^\frac{1}{3} M_\odot^\frac{2}{3}}
{kT_2} \left( \frac{M}{M_\odot} \right)^{\frac{2}{3}} E_{th} \simeq 10^{-2}
\alpha^{\frac{1}{3}}
\left( \frac{M}{M_\odot} \right)^{\frac{2}{3}} E_{th} \qquad\qquad (85)
$$

which yields for the upper limit masses $M \simeq 10^{12} M_\odot$

\markright{\centerline{-- \arabic{page} --}\hspace*{30cm}}

$$
| E_G | \simeq 10^6 E_{th} \qquad\qquad (86)
$$
Hence we conclude that the semimajor axis of the upper limit clouds decrease in a
rough approximation like a freely falling particle according to
$$
\ddot{r} (t) = - \frac{GM}{r^2(t)} \qquad\qquad(87 a)
$$
with the initial conditions
$$
{r} (0) = r_1 \, , \, \dot{r} (0) = 0 \qquad\qquad \qquad(87 b)
$$

Here we use the initial value $r(0) = r_1$ with $r_1$ according to eqs. (63) and
(71) since the mass outside of $r_1$ is negligible. The solution of eq. (87) reads
$$
\sqrt{2GM} \, t = r_1^\frac{3}{2} \left[ \mbox{ arctg } 
\sqrt{\frac{r_1}{r} - 1} + \sqrt{\frac{r}{r_1}}
\sqrt{1 - \frac{r}{r_1}}\, \right ] \qquad\qquad(88)
$$

With increasing contraction the cloud begins to rotate so that the contraction
will be retarded until it stops at a final radius $r_2$. We approximate the
contraction time $t_{contr}$ by
$$
t_{contr} = 2t_f \qquad\qquad\qquad(89)
$$

where $t_f$ is the free fall time (88) when $r = \frac{1}{2} (r_1 + r_2).$

We calculate the contraction times explicitly in an universe with $\varrho_0 =
10^{-30} g cm^{-3} (\alpha = 1)$ for b = 10 (see table 9). For $C = C_{min}$ we
have $M = 6,6 \cdot 10^{11} M_\odot$ which might be our galaxy including dark
matter. We obtain the initial value $r_1$ from eq. (73) by replacing $\mu_1$
by $10^{-5}$ (see eq. (71)). We find
$$
r_1 = 6,0 \cdot 10^5 ly \qquad \qquad(90)
$$
For the final value $r_2$ we take the present day value of our galaxy, hence
$$
r_2 = 0,5 \cdot{10}^5 ly \qquad\qquad (91)
$$
Then we obtain a contraction time
$$
t_{contr} = 2,5 \cdot 10^9 y \qquad\qquad(92)
$$
Calculating the red shift
$$
z = \frac{R_0}{R} - 1 \qquad\qquad\quad(93)
$$
corresponding to the time (92) from eq. (17) by use of a Hubble-parameter h = 70
we find
$$
z \simeq 3 \qquad\qquad (94)
$$
Let us consider the same galaxy at the earlier time $t = 0,5 t_{contr}.$ At this
time the radial flow velocities $v$ of matter are extremely high. From eq. (87) we
obtain
$$
v (r) = \sqrt{2GM} \sqrt{\frac{1}{r} - \frac{1}{r_1}} \qquad\qquad(95)
$$
which for
$r = \frac{1}{2} (r_1 + r_2) (t = t_f = 0,5 t_{contr})$ yields

$$
v \simeq 10^7 cm/sec \qquad \qquad(96)
$$
Comparing this velocity with the velocity of sound
$$
c_s = \left(\frac{kT}{m_p} \right)^\frac{1}{2}\qquad \qquad\qquad (97)
$$
for $T \simeq 300 K$ we find
\markright{\centerline{-- \arabic{page} --}\hspace*{30cm}}
$$
v \simeq 10^2 c_s \qquad\qquad\qquad\quad (98)
$$

Hence at the earlier time $t = 0,5 t_{contr}$ the galaxy is in a highly active
state where the redshift corresponding to this time is

$$
z \simeq 5, 8 \qquad \qquad\qquad\qquad (99)
$$
Next we consider in table 9 the parameter $C = 10^{-2}$
with a mass $M = 10^7 M_\odot$ which is the upper limit of globular clusters.
We have the ratio
$$
\frac{z_0}{r_0} = \frac{1}{14} \qquad\qquad\qquad\qquad (100)
$$
and from eq. (85)
$$ | E_G | \simeq 500 E_{th} \qquad \qquad\qquad(101) $$

We calculate the contraction time as above where we assume that the contraction
stops when the mass has become a sphere. Then we obtain
$$
t_{contr} = 1,6 \cdot 10^8 y \qquad\qquad\qquad (102)
$$
However, doing the same for $C = 0,9999$ with again
$M \simeq 10^7 M_\odot$ we find
$$
\frac{z_0}{r_0} = 71 \qquad\qquad\qquad\qquad(103)
$$
and a contraction time
$$
t_{contr} = 3 \cdot 10^9 y \qquad\qquad\qquad(104)
$$
Finally we obtain for $C = 0,999999$ with a mass $M \simeq 10^8 M_\odot$ the
ratio
$$
\frac{z_0}{r_0} = 707 \qquad\qquad (105)
$$
and the enormous contraction time
$$
t_{contr} = 30 \cdot 10^9 y \qquad\qquad (106)
$$

6) \underline{\bf Conclusion}

Obviously our results agree very well with observations. The upper limit masses
$M \simeq 10^{12} M_\odot$ evolve to fastly rotating spiral galaxies and are
ready for the evolution of stars about $2,5 \cdot 10^9 y$ after the big bang.
They are observed in their final state with a redshift $z \simeq 3$ and in
an earlier highly active state with a redshift $z \simeq 5$. The lower limit
masses $M \simeq 10^5 M_\odot$ are born as spheres and are therefore ready for
the evolution of stars a few $10^6$ years after the big bang. Hence the stars
in globular clusters are very old in agreement with observations.

What about the cigar--like clouds for $C \rightarrow 1 ?$ Maybe they have never
existed. However, provided they have existed their contraction time is too long
to evolve into  clouds which are more or less spherically symmetric. They rather
become unstable and decay into irregular clouds. Are these clouds the dark matter?

Finally we discuss the fact that the masses of the clouds do not depend on the
central value $\tau_c$ in the range (55) (see section 4.3). First of all we note
that the anisotropies must be \underline{axially symmetric} only in the
\underline{very inner region} since the density of the clouds falls practically
to zero when the variable $y = \tau/\tau_c$ drops to $y = 0,99$ (see eq. (59 b)).
Hence an anisotropy $\tau_c \simeq 10^{-5}$ can be composed of
say 100 axially symmetric anisotropies $\tau_c \simeq 10^{-7}$ and covers then
about 100 galaxies. Thus the spatial pattern of anisotropies of the background
radiation after recombination might have been composed of chains and slices and
hence matter has been arranged in chains and slices of galaxies. To this initial
pattern of galaxies the machinery used by the Virgo Consortium [2] can be applied
to give the large scale structures we observe today. However, in contrast to the
results by the Virgo Consortium the structures consist now of \underline{luminous}
matter.
\markright{\centerline{-- \arabic{page} --}\hspace*{30cm}}
\newpage
\begin{center}{\large{\bf REFERENCES}}
\end{center}
\begin{tabular}{ll}
[1] & S. Weinberg, Gravitation and Cosmology,\\
    & J. Wiley, New York (1972)\\[1ex]
[2] & Virgo Consortium (F. Pearce, P. Thomas, A. Jenkins,\\
    & C. S. Frenk, H. Couchman, S. White, J. Colberg,\\
    & G. Efstathiou, A. Nelson, J. Peacock),\\
    & Annual Report 1995, Max--Planck--Institute Garching\\[1ex]
[3] & G. Lessner, GRG 26 (1994), 385\\[1ex]
[4] & G. Lessner, GRG 27 (1995), 417\\[1ex]
[5] & G. Lessner, Astrophys. and Space Sci. 161 (1989), 175\\[1ex]
[6] & J. M. Stuart, Cargèse Lectures in Physics 6,\\
    & E. Schatzman, ed., Gordon and Breach, New York (1973)\\[1ex]
[7] & W. Israel, in General Relativity, L. O'Raifeartaigh, ed.\\
    & Clarendon Press, Oxford (1972)\\[1ex]
[8] & H. Dehnen, O. Obregon, Astron. and Astrophys. 12 (1971), 161\\[1ex]
[9] & B. Chaboyer, Y.C. Kim, Ap. J. 454 (1995), 767\\[1ex]
[10]& M. Salaris, S. Degl'Innocenti, A. Weiss, Ap. J. 479 (1997), 665\\[1ex]
[11]& S. Chapman, T.G. Cowling, The Mathematical Theory of\\
    & Non--Uniform Gases, Cambridge University Press, Cambridge (1970)\\[1ex]
[12]& C. Cercignani, Theory and Application of the Boltzmann Equation,\\
    & Scottish Academic Press, Edinburgh, London (1975)\\[1ex]
[13]& C.Marle, Ann. Inst. H. Poincare 10 (1969), 67\\[1ex]
[14]& J. D. Nightingale, Ap. J. 185 (1973), 105\\[1ex]
[15]& J. L. Anderson, H. R. Witting, Physica 74 (1974), 466\\[1ex]
[16]& J. L. Anderson, A. C. Payne, Physica 85 A (1976), 261\\[1ex]
[17]& R. Dominguez--Tenreiro, R. Hakim, Phys. Rev. D 15 (1977), 1435\\[1ex]
[18]& A. Majorana, J. Math. Phys. 31 (1990), 2042\\[1ex]
[19]& F. P. Wolvaardt, R. Maartens, Class. Quant. Grav. 11 (1994), 203;\\
    & 14(1997), 535\\[1ex]
\end{tabular}
\markright{\centerline{-- \arabic{page} --}\hspace*{30cm}}
\newpage

\markright{\centerline{-- \arabic{page} --}\hspace*{30cm}}
\huge{\begin{tabular}{c|c|c|c|c}
$\alpha$ & 0.5 & 1 & 2 & 5\\ \hline
$t_1 [10^6 y]$ & 3,7 & 2,6 & 1,9 & 1,2 \\ \hline
$\Delta t [10^6 y]$ & 2,9 & 2,1 & 1,5 & 0,92 \\ \hline
$\mu_1$ & 1,8 $\cdot 10^{-9}$ & 1,3 $\cdot 10^{-9}$ & 9,2 $\cdot 10^{-10}$ &
 5,8 $\cdot 10^{-10}$ \\ \hline
$\frac{1}{2} c \Delta t / R_1$ & 0,04 & 0,03 & 0,02 & 0,01 \\
\end{tabular}}

\vspace*{2,5cm}

\huge{ Table 1: The values of $ t_1, \Delta t $ and $ \mu_1 $ \\
\hspace*{2,5cm} for b = 10 $ ( T_1 = 1,08 \cdot 10^3 K) $ }

\vspace*{2,5cm}

\huge{\begin{tabular}{c|c|c|c|c}
$\alpha$ & 0.5 & 1 & 2 & 5\\ \hline
$t_1 [10^6 y]$ & 5,9 & 4,2 & 3,0 & 1,9 \\ \hline
$\Delta t [10^6 y]$ & 5,1 & 3,6 & 2,6 & 1,6\\ \hline
$\mu_1$ & 6,7 $\cdot 10^{-10}$ & 4,8 $\cdot 10^{-10}$ & 
$3,4 \cdot 10^{-10}$ & 2,1 $\cdot 10^{-10}$ \\  
\end{tabular}}

\vspace*{2,5cm}

\huge{ Table 2: The values of $t_1, \Delta t$ and $\mu_1 $  \\
\hspace*{2,5cm} for b = 20 ($ T_1 = 0,792 \cdot 10^3 K $)}

\newpage
\markright{\centerline{-- \arabic{page} --}\hspace*{30cm}}

\huge{\begin{tabular}{c|c|c|c|c}
$\,\alpha$ & 0.5 & 1 & 2 & 5\\ \hline
$\, M/M_\odot$ & 2,7 $\cdot 10^5$ & 2,5 $\cdot 10^5$
& 2,4 $\cdot 10^5$ & 2,2 $\cdot 10^5 $ \\ \hline
D [ ly ]& 870 & 820 & 770 & 700\\
\end{tabular}}

\vspace*{2,5cm}

\huge{ Table 3: Masses M and diameters D of gas\\
\hspace*{2,5cm} clouds in the centre if spherically\\ 
\hspace*{2,5cm}symmetric anisotropies for b = 10\\
\hspace*{2,5cm}$(T_1 = 1,08 \cdot 10^3 K, q = 0,0083)$

\vspace*{2,5cm}}

\huge{\begin{tabular}{c|c|c|c|c}
$\, \alpha$ & 0.5 & 1 & 2 & 5\\ \hline
$\, M/M_\odot$ & 2,8 $\cdot 10^5$ & 2,6 $\cdot 10^5$
& 2,5 $\cdot 10^5 $ & 2,4 $\cdot 10^5$ \\ \hline
D [ ly ]& 1240 & 1160 & 1070 & 960\\ 
\end{tabular}}

\vspace*{2cm}

\huge{Table 4: Masses M and diameters D of gas\\
\hspace*{2,5cm} clouds in the centre of sherically\\
\hspace*{2,5cm} symmetric anisotropies for b = 20\\
\hspace*{2,5cm} $(T_1 = 0,792 \cdot 10^3 K, q = 0,0024)$ }

\vspace*{2cm}
\newpage
\markright{\centerline{-- \arabic{page} --}\hspace*{30cm}}

\huge{\begin{tabular}{c|c|c|c|c}
$\, \alpha$ & 0.5 & 1 & 2 & 5\\ \hline
$\varepsilon_1$ & 6 $\cdot 10^{-16}$ & 2 $\cdot 10^{-24}$ & $ 10^{-38}$ &
$10^{-71}$ \\ \hline
$\mu_1$ & 1,8 $\cdot 10^{-9}$ & 1,3 $\cdot 10^{-9}$ & 
9,2 $\cdot 10^{-10}$ & 5,8 $\cdot 10^{-10}$ 
\end{tabular}}

\vspace*{2cm}

\huge{Table 5: The parameter $\varepsilon_1$ for b = 10\\
\hspace*{2,5cm}$(T_1 = 1,08 \cdot 10^3 K,$ q = 0,0083)}

\vspace*{2cm}

\huge{\begin{tabular}{c|c|c|c|c}
$\, \alpha$ & 0.5 & 1 & 2 & 5\\ \hline
$\varepsilon_1$ & $10^{-35}$ & $10^{-56}$ & $ 5 \cdot 10^{-90}$ &
$10^{-164}$ \\ \hline
$\mu_1$ & 6,7 $\cdot 10^{-10}$ & 4,8 $\cdot 10^{-10}$ & 3,4 
$\cdot 10^{-10}$ & 2,1 $\cdot 10^{-10}$
\end{tabular}}

\vspace*{2cm}

\huge{Table 6: The parameter $\varepsilon_1$ for b = 20\\
\hspace*{2,5cm}($T_1 = 0,792 \cdot 10^3 K,$ q = 0,0024)}

\newpage
\markright{\centerline{-- \arabic{page} --}\hspace*{30cm}}

\huge{\begin{tabular}{c|c|c|c|c}
$\, \alpha$ & 0.5 & 1 & 2 & 5\\ \hline
$C_{\mbox{min}}$ & $1,21 \cdot 10^{-7}$ & $1,50 \cdot 10^{-7}$ & 
$1,87 \cdot 10^{-7}$ & $2,75 \cdot 10^{-7}$ \\ \hline
$C_{\mbox{max}}$ & 0,999999 & 0,999999 & 0,999999 & 0,999999\\
&\qquad \,939 & \qquad \,925 & \qquad \, 906 & \qquad \,862
\end{tabular}}  

\vspace*{2cm}

\huge{Table 7: The values of $C_{\mbox{min}}$\\
\hspace*{2,5cm}and $C_{\mbox{max}}$ for b = 10\\
\hspace*{2,5cm} $(T_1 = 1,08 \cdot 10^3$ K, q = 0,0083)}
 
\vspace*{2,5cm}

\huge{\begin{tabular}{c|c|c|c|c}
$\, \alpha$ & 0.5 & 1 & 2 & 5\\ \hline
$C_{\mbox{min}}$ & $4,66 \cdot 10^{-8}$ & $5,95 \cdot 10^{-8}$ & 
$7,34 \cdot 10^{-8}$ & $1,10 \cdot 10^{-7}$ \\ \hline
$C_{\mbox max}$ & 0,999999 & 0,999999 & 0,999999 & 0,999999\\
&\qquad \,976 & \qquad \,970 & \qquad \, 963 & \qquad \,945
\end{tabular}} 
 
\vspace*{2,5cm}

\huge{Table 8: The values of $C_{\mbox{min}}$\\
\hspace*{2,5cm}and $C_{\mbox{max}}$ for b = 20\\
\hspace*{2,5cm} $(T_1 = 0,792 \cdot 10^3$ K, q = 0,0024)}

\begin{landscape}
\markright{\centerline{-- \arabic{page} --}\hspace*{30cm}}

\LARGE{\begin{tabular}{c|c|c|c|c}
$C$ & $M/M_\odot (\alpha = 0,5)$ & $M/M_\odot (\alpha = 1)$ & 
$M/M_\odot (\alpha = 2)$  & $M/M_\odot (\alpha = 5)$\\ \hline
$C_{min}$ & 8,5 $\cdot 10^{11}$ & 6,6 $\cdot 10^{11}$ & 4,9 $\cdot 10^{11}$ &
3,1 $\cdot 10^{11}$\\
$10^{-6}$ & 1.0 $\cdot 10^{11}$ & 9,8 $\cdot 10^{10}$ & 9,1 $\cdot 10^{10}$ & 
8,5 $\cdot 10^{10}$\\
$10^{-5}$ & 1.0 $\cdot 10^{10}$ & 9,8 $\cdot 10^{9}$ & 9,1 $\cdot 10^{9}$ & 
8,5 $\cdot 10^{9}$\\
$10^{-4}$ & 1.0 $\cdot 10^{9}$ & 9,8 $\cdot 10^{8}$ & 9,1 $\cdot 10^{8}$ & 
8,5 $\cdot 10^{8}$\\
$10^{-3}$ & 1.0 $\cdot 10^{8}$ & 9,8 $\cdot 10^{7}$ & 9,1 $\cdot 10^{7}$ & 
8,5 $\cdot 10^{7}$\\
$10^{-2}$ & 1.0 $\cdot 10^{7}$ & 9,8 $\cdot 10^{6}$ & 9,2 $\cdot 10^{6}$ & 
8,6 $\cdot 10^{6}$\\
$0,1$ & 1.0 $\cdot 10^{6}$ & 1,0 $\cdot 10^{6}$ & 9,6 $\cdot 10^{5}$ & 
9,0 $\cdot 10^{5}$\\
$2/3$ & 2,7 $\cdot 10^{5}$ & 2,5 $\cdot 10^{5}$ & 2,4 $\cdot 10^{5}$ & 
2,2 $\cdot 10^{5}$\\
$0,9$ & 3,6 $\cdot 10^{5}$ & 3,4 $\cdot 10^{5}$ & 3,2 $\cdot 10^{5}$ & 
3,0 $\cdot 10^{5}$\\
$0,99$ & 1.0 $\cdot 10^{6}$ & 9,9 $\cdot 10^{5}$ & 9,2 $\cdot 10^{5}$ & 
8,6 $\cdot 10^{5}$\\
$0,9999$ & 1.0 $\cdot 10^{7}$ & 9,8 $\cdot 10^{6}$ & 9,1 $\cdot 10^{6}$ & 
8,5 $\cdot 10^{6}$\\
$0,999999$ & 1.0 $\cdot 10^{8}$ & 9,8 $\cdot 10^{7}$ & 9,1 $\cdot 10^{7}$ & 
8,5 $\cdot 10^{7}$\\
$C_{max}$ & 4.2 $\cdot 10^{8}$ & 3,6 $\cdot 10^{8}$ & 3,0 $\cdot 10^{8}$ & 
2,3 $\cdot 10^{8}$\\
\end{tabular}}

\vspace*{1.5cm}

\LARGE{Table 9: Masses of gas clouds in the centre\\
\hspace*{2,5cm}of axially symmetric anisotropies\\
\hspace*{2,5cm}for b = 10  $(T_1 = 1,08 \cdot 10^3$ K, q = 0,0083)}

\newpage
\markright{\centerline{-- \arabic{page} --}\hspace*{30cm}}

\LARGE{\begin{tabular}{c|c|c|c|c}
$C$ & $M/M_\odot(\alpha = 0,5)$ & $M/M_\odot(\alpha = 1)$ & 
$M/M_\odot(\alpha = 2)$  & $M/M_\odot(\alpha = 5)$\\ \hline
$C_{min}$ & 2,3 $\cdot 10^{12}$ & 1,7 $\cdot 10^{12}$ & 1,3 $\cdot 10^{12}$ &
8,6 $\cdot 10^{11}$\\
$10^{-7}$ & 1.1 $\cdot 10^{12}$ & 1,0 $\cdot 10^{12}$ & 9,7 $\cdot 10^{11}$ & 
\\
$10^{-6}$ & 1.1 $\cdot 10^{11}$ & 1,0 $\cdot 10^{11}$ & 9,7 $\cdot 10^{10}$ & 
9,4 $\cdot 10^{10}$\\
$10^{-5}$ & 1.1 $\cdot 10^{10}$ & 1,0 $\cdot 10^{10}$ & 9,7 $\cdot 10^{9}$ & 
9,4 $\cdot 10^{9}$\\
$10^{-4}$ & 1.1 $\cdot 10^{9}$ & 1,0 $\cdot 10^{9}$ & 9,7 $\cdot 10^{8}$ & 
9,4 $\cdot 10^{8}$\\
$10^{-3}$ & 1.1 $\cdot 10^{8}$ & 1,0 $\cdot 10^{8}$ & 9,7 $\cdot 10^{7}$ & 
9,4 $\cdot 10^{7}$\\
$10^{-2}$ & 1.1 $\cdot 10^{7}$ & 1,0 $\cdot 10^{7}$ & 9,8 $\cdot 10^{6}$ & 
9,5 $\cdot 10^{6}$\\
$0,1$ & 1.1 $\cdot 10^{6}$ & 1,1 $\cdot 10^{6}$ & 1,0 $\cdot 10^{6}$ & 
9,9 $\cdot 10^{5}$\\
$2/3$ & 2,8 $\cdot 10^{5}$ & 2,6 $\cdot 10^{5}$ & 2,5 $\cdot 10^{5}$ & 
2,5 $\cdot 10^{5}$\\
$0,9$ & 3,8 $\cdot 10^{5}$ & 3,6 $\cdot 10^{5}$ & 3,4 $\cdot 10^{5}$ & 
3,3 $\cdot 10^{5}$\\
$0,99$ & 1.1 $\cdot 10^{6}$ & 1,0 $\cdot 10^{6}$ & 9,8 $\cdot 10^{5}$ & 
9,5 $\cdot 10^{5}$\\
$0,9999$ & 1.1 $\cdot 10^{7}$ & 1,0 $\cdot 10^{7}$ & 9,7 $\cdot 10^{6}$ & 
8,5 $\cdot 10^{6}$\\
$0,999999$ & 1.0 $\cdot 10^{8}$ & 1,0 $\cdot 10^{8}$ & 9,7 $\cdot 10^{7}$ & 
9,4 $\cdot 10^{6}$\\
$C_{max}$ & 7.0 $\cdot 10^{8}$ & 5,9 $\cdot 10^{8}$ & 5,1 $\cdot 10^{8}$ & 
4,0 $\cdot 10^{8}$\\
\end{tabular}}

\vspace*{1.5cm}

\LARGE{Table 10: Masses of gas clouds in the centre \\
\hspace*{3cm}of axially symmetric anisotropies \\
\hspace*{3cm}for b = 20  $(T_1 = 0,792 \cdot 10^3$ K, q = 0,0024) }

\end{landscape}

\end{document}